\documentclass{article}
\usepackage{spconf,amsmath,amsfonts, graphicx, float}
\usepackage{multirow}
\usepackage{makecell}
\usepackage{color}
\usepackage{booktabs}
\DeclareMathOperator*{\argmax}{argmax}
\DeclareMathOperator*{\concat}{Concat}

\title{Minimum word error training for  non-autoregressive Transformer-based code-switching ASR}
\name{Yizhou Peng$^{1*}$, Jicheng Zhang$^{1*}$, Haihua Xu$^{2*}$, Hao Huang$^1$, Eng Siong Chng$^3$\thanks{$*$ authors have equal contributions. Hao huang is the correspondence author. }}
\address{$^1$School of Information Science and Engineering, Xinjiang University, China \\  $^2$Bytedance AI Lab, Singapore\\
   $^3$School of Computer Science and Engineering, Nanyang Technological University, Singapore
}
\begin{document}
\ninept
\maketitle
\begin{abstract}
% The abstract should appear at the top of the left-hand column of text, about
 Non-autoregressive end-to-end ASR framework might be potentially appropriate for code-switching recognition task thanks to its inherent property that  present output token being independent of historical ones. However, it still under-performs the state-of-the-art autoregressive ASR frameworks. In this paper, we propose various approaches to boosting the performance of a CTC-mask-based non-autoregressive 
Transformer under code-switching ASR scenario. To begin with, we attempt diversified masking method that are closely related with code-switching point, yielding an improved baseline model. More importantly,
 we
 employ Minimum Word Error (MWE) criterion to train the model. 
 One of the challenges is how to generate a diversified hypothetical space, so as to obtain the average loss for a given ground truth. % is this sentence needed here?
 To address such a challenge, we explore different approaches to
 % to randomize the mask of CTC output for a given utterance, 
 yielding  desired N-best-based hypothetical space. We demonstrate the efficacy of the proposed methods on SEAME corpus, a challenging English-Mandarin code-switching corpus for Southeast Asia community. Compared with the cross-entropy-trained strong baseline, the proposed MWE training method achieves consistent % consistent
 performance improvement on the test sets. 
\end{abstract}
\section{Introduction}
\label{sec:intro}
End-to-end~(E2E) ASR has become predominant in both research and industrial areas thanks to its modeling simplicity, compactness, as well as efficacy in performance in recent years. The state-of-the-art~(SOTA) E2E ASR modeling frameworks include frame-synchronous family, such as Connectionist Temporal Classification~(CTC)~\cite{ICML-CTC-2006,ICASSP-CTC-2018}, Recurrent Neural Network Transducer (RNN-T)~\cite{RNN-T,ASRU-RNNT-2017}, and lab-synchronous or attention-based encoder-decoder (AED) family,  such as Transformer~\cite{ICASSP-transformer-2018,INTERSPEECH-transformer-2021,vaswani2017attention}, and Conformer~\cite{INTERSPEECH-conformer-2020} etc. Except for the CTC method, all these belong to auto-regressive % AT means auto-regressive transformer, NAT means non-auto-regressive transformer 
modeling framework, that is, the present output token of the decoder is dependent on previously predicted tokens. 

The autoregressive method has one obvious   advantage that it can simultaneously learn acoustic and language semantics jointly, yielding SOTA results. However, one of the limitations of the autoregressive method is that it is weak in recognition of unseen utterances. Take code-switching ASR as an example, we cannot let an autoregressive-based ASR system that is learned with two mixed monolingual data recognize code-switching words in a given utterance~\cite{INTERSPEECH-constraint-2019}.

More recently, non-autoregressive Transformer (NAT)  modeling framework has drawn growing attention in machine translation~\cite{arXiv-2018-nonautoregressive,EMNLP-NonAutoregressiveMT-2020}, as well as speech recognition~\cite{arXiv-nat-2020,ng21b_interspeech,guo21_interspeech,song21_icassp_nat} societies due to its output token from the decoder can be independent to each other, resulting in faster decoding while comparable results to what is brought by the autoregressive Transformer (AT). However, the performance of the NAT method is still worse than that of its AT counterparts.

In this paper, we attempt to employ NAT
% ~\cite{arXiv-nat-2020}
for code-switching speech recognition. We are motivated by that the output token independence property of the NAT method could be potentially more appropriate for code-switching ASR given more monolingual data available. However, our first priority here is still to make the NAT performance approach that of the corresponding AT method. Consequently, we are meant to boost the SOTA non-autoregressive performance  under code-switching ASR scenario. 

Employing the CTC-mask-based NAT, our efforts lie in the following aspects. 1) Expedite the training process by an enforced Conditional Masked Language Model (CMLM) learning. 2) We employ different masking methods that aim at code-switching points in utterance for training, yielding improved baseline results. 3) Furthermore, we propose to employ Minimum Word Error (MWE) criterion to train our NAT model, yielding consistent performance improvement. To the best of knowledge, we are the first to introduce MWE criterion to train the NAT model for speech recognition task.

\section{Related works} % Needed?
\label{sec:related-works}
Recently NAT has been widely studied in ASR area due to its much faster speed in inference
while it can achieve comparable recognition results, in contrast to its  AT counterpart.
Though with the same NAT framework, people employ it to do the ASR work differently. 
One school of branch advocates using one-pass inference method, that is, once the embedding of the encoder representation for a given utterance is ready, one can  let the decoder output the overall inference sequence with one-pass simultaneously~\cite{spike-triggered-2020}. Another school of  branch attempts to output the inference results with iterative fashion, where the decoder input is a sequence of which some of tokens are randomly masked, and the inference process is the process of predicting those masked tokens~\cite{MaskCTC,chuang2021nonautoregressive,MaskCTC-conformer}. In whatever approach is taken, the length of the token sequence that is fed into the decoder is essential for the inference performance. %~\cite{}.
In this paper, we take the iterative inference approach, following what is proposed in~\cite{MaskCTC,lee-etal-2018-deterministic}. Our efforts are focused on extending the prior work under the code-switching speech recognition environment.  More importantly, we attempt to employ MWE criterion to train the NAT network, yielding consistent improvement.

\section{System Framework}~\label{sec:sys-frame}
Theoretically, given an input sequence $\bf{X}=\{{\bf{x}_1, ..., \bf{x}_T}\}$ for both AT and NAT, 
the objective is to predict the conditional probability $P(y_i|\bf{X}; \theta)$ of each token $y_i$, where $y_i \in {\bf{Y}}=\{y_1,..., y_N\}$. 
The objective of AT and NAT  ASR is the same, but it is accomplished differently. 

%AT proceeds step-by-step for inference, while NAT can output the entire sequence $\bf{Y}$ simultaneously for the one-pass method, and constant $K$ steps, where $K \ll\lvert N$ for the iterative method.   

Technically, the whole NAT network is made up of 3 sub-networks, i.e., Transformer encoder, CTC network on top of the Transformer encoder, and a self-attention based Transformer decoder. For AT model, CTC is optional though it can benefit the AT's performance~\cite{ICASSP-CTC-ATT-2017, Hybrid-CTC-ATT-2017}. For NAT model, CTC is critical in both one-pass and iterative decoding fashions~\cite{spike-triggered-2020,MaskCTC,chuang2021nonautoregressive,MaskCTC-conformer}, except that \cite{Listen-Attentively-2020} directly employs a fixed length position encoded sequence as input to the decoder, producing sub-optimal results. In our iterative NAT method, CTC is employed to generate preliminary inference sequence by greedy method, besides, it is shown in~\cite{MaskCTC}, CTC training is essential to producing a desired performance. 

For clarification to what follows, we start with revisiting the principle of working mechanism for  AT, then  CTC, and finally our  Conditional Mask Language Model~(CMLM)-based~\cite{CMLM-2019} NAT in this section.
\subsection{Autoregressive Transformer}~\label{sub:at}
For Transformer, the joint probability is estimated as follows:
\begin{equation}
P_{\text{AT}}({\bf{Y}|\bf{X}}) = \prod_{i=1}^{N}P(y_i|\bf{y}_{<i}, X)
\label{eq:at}
\end{equation}
where $\bf{y}_{<i}$ refers to the previous tokens before $i$.
During training, estimating $P(y_i|\bf{y}_{<i}, X; \theta)$ can proceed with parallelism since $\bf{y}_{<i}$ is directly from ground truth, while the inference process has to proceeds  step-by-step, as present output token $y_i$ depends on $\bf{y}_{<i}$. 
\subsection{Connectionist Temporal Classification}~\label{sub:ctc}
For CTC, the joint probability is estimated as follows:
\begin{equation}
P({\bf{Y}^{\prime}|X}) = \prod_{t=1}^{T}P(y_t^{\prime}|\bf{X})
\label{eq:ctc-token}
\end{equation}
where $y_t^{\prime} \in \bf{Y}^{\prime}$ is a CTC frame-synchronous token that is either a normal token as in Section~\ref{sub:at}, or a blank token ``-", indicating nothing being output. 
For training,  the objective function is 
\begin{equation}
P_{\text{CTC}}(\bf{Y|X}) = \sum_{\mathcal{T}^{-1}(\bf{Y}^{\prime}) = \bf{Y}}  P(\bf{Y}^{\prime}|X)   \label{eq:ctc-train}
\end{equation}
where function $\mathcal{T}^{-1}$ removes all duplicated and blank tokens. For inference, CTC can choose either beam search based forward-search algorithm, or greedy algorithm. The forward-search algorithm is governed by the following formulae:
\begin{equation}
{\bf{Y}^{*}} = \argmax_{\mathcal{T}^{-1}(\bf{Y}^{\prime}) = \bf{Y^{*}}} P(\bf{Y^{\prime}|X}) \label{eq:ctc-inf}  
\end{equation}
Normally, the forward-search algorithm is  time consuming. In this paper, to preserve the NAT faster inference property, we employ the greedy algorithm  as follows instead:
\begin{align}
    &{\bf Y}^{\prime*} = \concat_{1\le t\le T}\argmax_{y_t^{\prime*}}
    P(y_t^{\prime}|\bf{X}) \label{eq:ctc-greed01}\\
    &{\bf Y^*}_{\text{greed}} = \mathcal{T}^{-1}({\bf Y}^{\prime*}) \label{eq:ctc-greed02}
\end{align}
though the greedy algorithm yields much worse results, they will be refined by the NAT decoder iteratively.

\subsection{CMLM non-autoregressive ASR}~\label{sub:nat}
As observed from Equation~(\ref{eq:ctc-token}), we know CTC belongs to non-autoregressive modeling framework by definition.
If we use the greedy method (see Equations (\ref{eq:ctc-greed01}) and (\ref{eq:ctc-greed02})) to perform  inference, CTC is completely a non-autoregressive ASR method, since each output is independent to the remaining ones. However, it yields much worse ASR performance. In this paper, inspired from AT method, as well as pursuing faster inference speed from NAT method, we employ a CMLM NAT ASR framework~\cite{MaskCTC} as follows:
\begin{equation}
P_{\text{CMLM}}({\bf{Y}}_{\text{mask}}|{\bf{Y}_{\text{obs}}, \bf{X}}) = \prod_{y \in \bf{Y}_{\text{mask}}} P(y|{\bf{Y}}_{\text{obs}}, \bf{X})~\label{eq:cmlm}
\end{equation}
where ${\bf Y}_{\text{obs}}= \bf{Y}~\backslash~{\bf Y}_{\text{mask}}$. That is, the CMLM NAT method attempts to predict mask tokens $\bf{Y}_{\text{mask}}$, conditioning on unmasked/observed tokens $\bf{Y}_{\text{obs}}$, given an acoustic input sequence $\bf{X}$ from encoder. 

During training, the ground truth tokens are randomly masked, and they are replaced with a predefined mask symbol \texttt{<MASK>}. As a result, the NAT decoder is learned as a CMLM. During inference, the output of CTC using Equation~(\ref{eq:ctc-greed02}) takes the place of the ground truth, but some of the tokens, which have lower posteriors, are similarly masked. An iterative inference method is employed to yield the final results.

In ~\cite{MaskCTC}, it is observed that using only criterion in Equation~(\ref{eq:cmlm}) to train the NAT model results in worse performance, and a joint CTC and CMLM criterion is employed instead as follows:
\begin{align}
     \mathcal{L}_{\text{NAT}} = \alpha P_{\text{CTC}}(\bf{Y|X})
      & +(1-\alpha) P_{\text{CMLM}}(\bf{Y_{\text{mask}}|Y_{\text{obs}}, X})\label{eq:nat-loss}
\end{align}
where $\alpha$ is  a hyperparameter fixed with 0.3 for all experiments in what follows. 
 
\section{Proposed Methods}~\label{sec:proposed}
\subsection{Enforced CMLM learning for training expedition}~\label{sub:expedite}
\begin{figure}[htbp]
    \centering
    \caption{Accuracy trend on both training and validation data sets, using the proposed enforced CMLM learning criterion.} 
    \label{fig:training-acc-curve}
    \includegraphics[width=8cm]{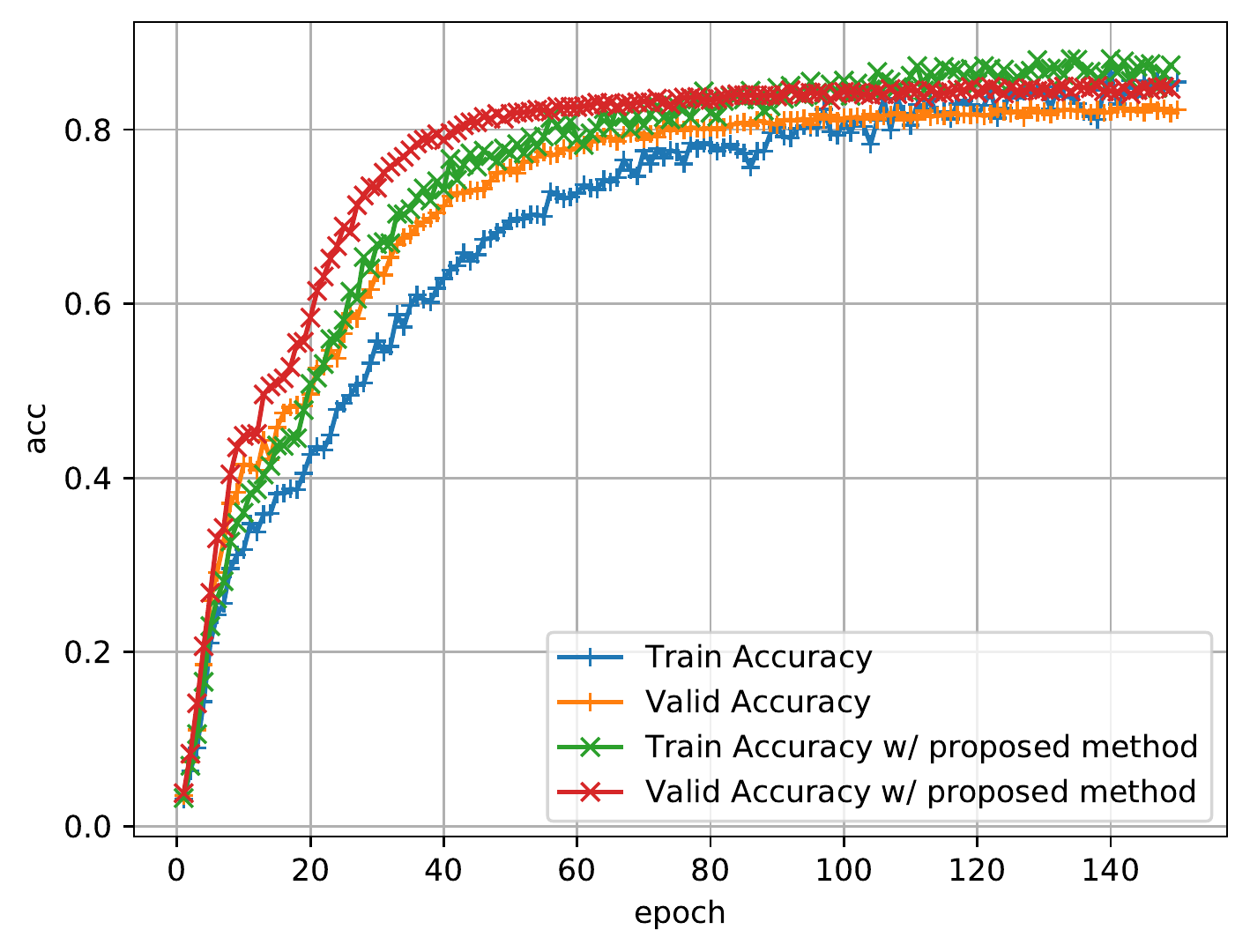}
\end{figure}

In our training practice with criterion~(\ref{eq:nat-loss}), we found the training is much slower to converge compared with the normal AT training. We guess the loss from the term $P_{\text{CMLM}}(\bf{Y_{\text{mask}}|Y_{\text{obs}}, X})$ cannot make the network sufficiently trained. Therefore, we propose to  add another term to make it faster to converge. Specifically, we not only employ the observed sequence to predict masked sequence, we also do the opposite prediction, that is, let the masked tokens unmasked while the observed tokens masked. 
Consequently, the criterion ~(\ref{eq:nat-loss}) is changed to
\begin{align}
     \mathcal{L^\prime}_{\text{NAT}} = \alpha P_{\text{CTC}}(\bf{Y|X})
       +(1-\alpha)[ &P_{\text{CMLM}}(\bf{Y_{\text{mask}}|Y_{\text{obs}}, X}) \nonumber \\
      + &P_{\text{CMLM}}(\bf{Y_{\overline{\text{mask}}}|Y_{\overline{\text{obs}}}, X}) ] \label{eq:nat-loss02}
\end{align}
Using the updated objective function~(\ref{eq:nat-loss02}), we find that it is faster to converge for training. Figure~\ref{fig:training-acc-curve} illustrates the accuracy curve on both training and validation data set using the proposed training method in contrast to the baseline. From Figure~\ref{fig:training-acc-curve}, we observe faster training convergence, as well as better accuracy on  both  training  and  validation  data while using the proposed training method.
For convenience, we name the original mask method as R-Mask method, while the proposed one as C-Mask method. 
% in figure~\ref{fig:example-text}, example of complimentary mask is at the end.

% For a given token list ${\bf{Y}}=\{y_1,..., y_N\}$, 
% The masking method mentioned in~\cite{MaskCTC} is a randomly mask of length ${\bf{L}\in[1, N]}$. For improving the robustness of the model, we propose a method to complement the mask, which is for a token sequence, first to randomly mask some of the tokens as mentioned above, then reverse the pair \texttt{<MASK>} and the corresponding token to generate a new masked sequence.

\subsection{Code-switching-based masking methods}~\label{sub:cs-masking}
Code-switching  refers to a multilingual phenomenon of which speakers spontaneously  switch between two or more different languages, and such switching can occur within an utterance, denoted as $\texttt{CS}_1$, or  between consecutive utterances, denoted as $\texttt{CS}_2$. Technically, $\texttt{CS}_1$ is much more challenging than $\texttt{CS}_2$, since cross-lingual word switching normally has rare samples, resulting in ASR models insufficiently learned and  weak in performing code-switching accurately. In what follows, we are dealing with $\texttt{CS}_1$ code-switching problem.

As the CMLM-based NAT ASR has a decoder that acts as a conditional masked language model, we can deliberately mask those words located at  code-switching points. Alternatively, we can also mask one language or another language. In either way, we are meant to enforce the CMLM to sufficiently learn code-switching predictions.

Specifically, given an English-Mandarin code-switching utterance ${\bf{Y}}=\{y_1,..., y_N\}$, we have two kinds of code-switching pairs, that is,
$\text{Pair}_1=(y_{\text{CN}}, y_{\text{EN}})$ or $\text{Pair}_2=(y_{\text{EN}}, y_{\text{CN}})$.  We are doing the following 4 types of masking method:
\begin{itemize}
    \item We mask the first token of those code-switching pairs, denoted as F-Mask.
    \item We mask the second token of the code-switching pairs, denoted as S-Mask.
    \item In consideration of language, we only mask the Mandarin tokens, denoted as M-Mask.
    \item Alternatively, we only mask the English tokens, denoted as E-Mask.
\end{itemize}

\subsection{Minimum word error training}~\label{sub:mwe}
% The goal of Minimum word error training is to minimize the expected word error rate. 
So far, CMLM network training criteria from Equations~(\ref{eq:nat-loss}) or (\ref{eq:nat-loss02}) belongs to the Cross Entropy (CE) category, i.e., computing the log likelihood of each ground truth posterior during  training. However, 
ASR performance metric is defined by calculating Word Error Rate (WER), an edit distance between the recognized sequence and ground truth, for each recognized utterance, that is, $\mathcal{D}(\bf{Y,Y^{*}})$, where $\bf Y^{*}$ and $\bf{Y}$ are ground truth and recognized sequences respectively. Therefore, a matched  
 training criterion for  any ASR system should be reducing the expected Word Error Rate,  $\mathbb{E}[\mathcal{D}(\bf{Y,Y^{*}})]$, on training data. From this perspective, the conventional CE training yields sub-optimal performance. In this section, we are exploring sequential-based Minimum Word Error (MWE)~\cite{MWER,INTERSPEECH-MWER,INTERSPEECH-neural-aligner} training under NAT framework. For simplicity, we employ N-best hypothesis list to approximate our hypothetical space. The training criterion is as follows:
 \begin{equation}
 \mathcal{L}_{\text{MWE}} = \mathbb{E}[\mathcal{D}({\bf{Y,Y^{*}}})] = \sum_{i=1}^{N}P(\bf{Y_i|X})\mathcal{D}(\bf{Y_i,Y^{*}})~\label{eq:mwe-loss}      
 \end{equation}
where $N$ is the number of hypothesis in the N-best list, $P(\bf Y_i|X)$ is the utterance posterior, calculated using Equation~(\ref{eq:ctc-token}) under NAT framework. 
In practice, to compute the MWE loss, we normalize both $P(\bf Y_i|X)$ and  $\mathcal{D}(\bf{Y_{i},Y^{*}})$ in Equation (\ref{eq:mwe-loss}) as advocated in~\cite{MWER}.
Besides, the final training criterion is an interpolation of CE and MWE as:
\begin{equation}
\mathcal{L^\prime}_{\text{MWE}} = \gamma \mathcal{L}_{\text{NAT}} + (1-\gamma) \mathcal{L}_{\text{MWE}} \label{eq:mwe-loss02}
\end{equation}
where $\gamma$ is a tunnable parameter and it is fixed with $0.01$.

The problem of using criterion (\ref{eq:mwe-loss02}) for training is how to generate N-best list under our NAT framework, which is quite simple under autoregressive framework since training proceeds with token-by-token fashion, and each token can have multiple choices under predefined beams. A simplified recipe in our case is to generate Nbest restricted to $\bf{Y}_{\text{mask}}$ subspace in Equation~(\ref{eq:nat-loss}), denoted as Output-Nbest. Alternatively, when we get output from CTC using Equation~(\ref{eq:ctc-greed02}), we can employ it to generate $N$ randomly masked sequences, yielding Nbest hypothesis. To differentiate, we call such an Nbest list generation method as Input-Nbest. In this paper, we employ both methods to test the efficacy of the MWE training criterion.

% \begin{figure}[htbp]
%     \centering 
%     \begin{tabular}{cc}
%         \begin{minipage}[t]{0.44\linewidth}
%             \includegraphics[width=1\linewidth]{N1-Best-new.pdf}
%             \centerline{(a)}\medskip
%         \end{minipage}
%         \begin{minipage}[t]{0.445\linewidth}
%             \includegraphics[width=1\linewidth]{NBest-new.pdf}
%             \centerline{(b)}\medskip
%         \end{minipage}
%     \end{tabular}
%     \caption{Two ways to extract N-Best Hypothesis}
%     \label{fig:Transfer-learning}
% \end{figure}

\section{Experiments}~\label{sec:exp}
\subsection{Data}~\label{sub:Data}
We conduct our experiments on SEAME data, which is a southeast English-Mandarin code-switching corpus, amounted to $\sim$110 hours in total. The data is collected from both Singapore and Malaysia areas. It is spontaneous, recorded under clean conversational environment. During conversing, one speaker is an interviewer, responsible for asking questions, and the questions are covered with diversified topics. The other speaker is an interviewee to answer the questions. SEAME corpus only contains interviewee's voice, and there are 154 speakers in total. 
One can refer to ~\cite{INTERSPEECH2018-SEAME} for more details.  
\begin{table}[htp]
 \centering
%  \caption{Statistics information for \texttt{Train}, \texttt{Valid},    $\texttt{eval}_{\texttt{man}}$ is dominated by Mandarin and ${eval}_{sge}$ is dominated by Singapore English.}
\caption{Main statistics for  data \texttt{Train}, \texttt{Valid},    $\texttt{eval}_{\texttt{man}}$, and $\texttt{eval}_{\texttt{sge}}$ respectively.}
 \label{tab:data-spec}
\begin{tabular}{ccccc}
 \toprule
    & \texttt{Train} & \texttt{Valid} &$\texttt{eval}_{\texttt{man}}$ &$\texttt{eval}_{\texttt{sge}}$ \\
   \midrule
    Duration (Hrs) & 93.6 & 7.0 & 7.5 & 3.9\\
    Ave. Length (Sec)  & 3.9 & 4.0 & 4.1 & 2.7 \\
    \#speaker & 134 & 134 & 10 & 10 \\
    
    \bottomrule
\end{tabular}
\end{table}

To evaluate ASR system's performance, we define 2 test sets, each contains the speech of 10 speakers' entire conversation. Each speaker is either from  Singapore or from Malaysia area.  The 2 data sets are named as  
$\texttt{eval}_{\texttt{man}}$ and $\texttt{eval}_{\texttt{sge}}$ respectively.
$\texttt{eval}_{\texttt{man}}$  is dominated with Mandarin, the English to Mandarin ratio is around 3:7 in terms to token counting. Here, token refers to English word and Mandarin character. By contrast, $\texttt{eval}_{\texttt{sge}}$ is dominated with English, and corresponding English to Mandarin ratio is around 7:3. 
Table~\ref{tab:data-spec} reports the overall data distribution for experiments.

\begin{table}[H]
 \centering
 \caption{MER (\%) for the proposed NAT methods in contrast with AT methods.  The SA stands for Spec-Augmentation. The R-Mask is for Random Mask. (EMFS)-Masks are  code-switching-based English Mask, Mandarin Mask, the First Mask and the Second Mask respectively, refer to Section~\ref{sub:cs-masking} for the details. For both Input-Nbest and Output-Nbest , refer to Section~\ref{sub:mwe} for details.}
 %First Mask, Chinese Mask, English Mask, and Second Mask respectively mentioned in~\ref{sub:cs-masking}.}
 \label{tab:maskctc-results}
\begin{tabular}{llccc}
\toprule
    \multicolumn{2}{c}{Modeling method \& Setup} & $\texttt{eval}_{\texttt{man}}$ & $\texttt{eval}_{\texttt{sge}}$ & RTF \\
    \midrule
    \multirow{2}*{\shortstack{AT+CE}} & 
    Transformer & 21.74 & 29.46 & 0.68 \\
    & \quad + SA & 17.89 & 25.1 & 0.70 \\
    \midrule
    \multirow{6}*{\shortstack{NAT + CE}} & R-Mask~\cite{MaskCTC} & 20.28 & 28.33 & 0.05 \\
        
    & C-Mask & 20.33 & 27.87 & 0.05 \\
    & E-Mask & 20.11 & 27.73 & 0.05 \\
    & M-Mask & \textbf{20.04} & 28.11 & 0.05 \\
    & S-Mask & 20.17 & 27.89 & 0.05 \\
    & F-Mask & 20.07 & \textbf{27.56} & 0.05 \\
    % \cline{2-5}
    \midrule
     \multirow{2}*{\shortstack{NAT+MWE}} & Output-Nbest & 19.97 & 27.34 & 0.05 \\
    & Input-Nbest & 19.90 & 27.29 & 0.05 \\
    \bottomrule
    
\end{tabular}
\end{table}
% The SEAME corpus is a spontaneous conversational bilingual speech corpus recorded by 154 speakers from Singapore and Malaysia using microphones~\cite{INTERSPEECH2010-SEAME,ICASSP2012-SEAME}.
% For all of our experiments, we follow the same data definition  as~\cite{INTERSPEECH2018-SEAME} except for that randomly extracting some utterances from training set for validation. Also, those utterances in training sets longer than 20 seconds are ignored for stabilizing the training procedure.
% Table~\ref{tab:data-spec} shows the data definition.
% % where ${eval}_{man}$ is dominated by Mandarin and ${eval}_{sge}$ is dominated by Singapore English.

\subsection{Modeling Setup}~\label{sub:models}  % Experimental Setup? Or Setup
All experiments are conducted with Transformer on Espnet toolkit~\cite{watanabe2018espnet}.
% are performed using E2E-based Transformer modeling framework with Espnet tookit~\cite{}.
% For all the transformer models, 
For Autoregressive Transformer (AT) models,
they are configured with 12-layer encoder, and 6-layer decoder with 4-head attention. The attention dimension is 256. 
The AT models are also jointly trained with CTC plus attention methods, of which the weighting factor for the CTC part is 0.3.
% We also apply hybrid CTC/attention multi-task learning method, with CTC weight of 0.3.
% For the autoregressive transformer model, it is configured with of encoder of 12 layers, decoder of 6 layers and 4 head attention. 
% For our proposed methods, all of the work is based on a CTC-mask-based NAT model proposed by~\cite{MaskCTC}. The model consists of a 12 layer transformer encoder, a CTC for frame-level alignment, and a 6 layer transformer decoder.
The input acoustic features are 83-dimensional including 80-dim log filter bank and 3-dim pitch features.  For better performance, we adopt 
data augmentation methods. They are speed perturbation~\cite{ko-2015-speed_perturb} and spec-augmentation~\cite{INTERSPEECH-SpecAug-2019} respectively.
The output of the Transformer is BPE set. It consists of two parts which are 3000 BPEs for English and 3920 for single Chinese characters from the training transcript.

For CE training,  all the Transformers are optimized with Noamoptimizer~\cite{vaswani2017attention}, with 0.1 drop-out configuration.
All the final models are obtained by averaging the model parameters of the last 5 epochs.  For MWE training, it proceeds with 5 epochs with  4-best hypothesis.
For AT decoding, We adopt beam-search-based decoding method, with beam size being 10. For the  CTC-mask-based NAT decoding, we follow the iterative inference fashion as in~\cite{MaskCTC}. The tokens whose CTC posteriors are below 0.9 are masked. Besides, we fix the max iterations with 10 during decoding.

% with the CTC posterior threshold  $0.9$ to mask, 

% we proceed with 0.1 drop-out, and all Transformers are optimized with   Besides, 
% For MWE training criterion, we train another 5 epochs based on the 50-th epoch of NAT model obtained by F-Mask method.
% For decoding, we  and \#Iteration=10 for Mask-based-CTC models and beamsize of 10 for AT models. 

\subsection{Performance metric}~\label{sub:metric}
Two main metrics are employed to evaluate ASR performance. They are MER(\%) and RTF respectively.
MER(\%) is an abbreviation for Mixed Error Rate that consists of word error rate for English and character error rate for Mandarin.
The Real Time Factor~(RTF) is obtained under the scenario where
all test sets are decoded with  single thread, by a single  core of Intel(R) Xeon(R) Gold 6238R, 2.20GHz.

% which is measured for decoding all evaluation sets with single core of , and all the models are obtained by averaging the model parameters of the of last 5 epochs.

\subsection{Results}~\label{sub:result}
Table~\ref{tab:maskctc-results} presents the MER (\%) and RTF results of the proposed NAT methods  on the two test sets in contrast with the AT method.
% Table~\ref{tab:maskctc-results} reports the results of the proposed NAT methods in contrast with AT methods for two evaluation sets on Mixed Error Rate (MER\%) and RTF. 
From table~\ref{tab:maskctc-results}, 
we observe that the proposed C-Mask method  outperforms the R-Mask baseline on $\texttt{eval}_{\texttt{sge}}$ with slightly degraded MER result on $\texttt{eval}_{\texttt{man}}$. 
For the code-switching-based masking methods, 
% we find that the improved results for each evaluation set are from its corresponding language masking method, 
it seems that language specific masking method favors corresponding language dominated test set,
that is, the E-Mask improves on $\texttt{eval}_{\texttt{sge}}$ while  the M-Mask obtains the best MER on $\texttt{eval}_{\texttt{man}}$. 
Among all the masking methods, F-Mask gets the best MER on $\texttt{eval}_{\texttt{sge}}$ and a second best MER on $\texttt{eval}_{\texttt{man}}$. So we continue our MWE training from F-Mask NAT model.
For the MWE training, both Nbest generation methods yield consistent performance improvement over the best NAT CE training method, and the Input-Nbest slightly outperforms the Output-Nbest.
% so we continue the MWE criterion training based on F-Mask model and get further improvements on both evaluation sets while applying either Output-Nbest or Input-Nbest setup, and Input-Nbest slightly outperforms Output-Nbest.
Finally, though all of the proposed methods for the NAT framework still have gap with the AT one in terms of MER, we still obtain 0.38\% and 1.04\%  absolute MER reduction  respectively compared with the baseline R-Mask method. Additionally, the RTF of the proposed NAT method is over 10x times faster.

\begin{table}[H]
 \centering
 \caption{MER(\%) of  words located at code-switching points, as defined in Section~\ref{sub:cs-masking}. Refer to Table~\ref{tab:maskctc-results} for the details of name notation.}
 %First Mask, Chinese Mask, English Mask, and Second Mask respectively mentioned in~\ref{sub:cs-masking}.}
 \label{tab:CS-WER-results}
\begin{tabular}{llcc}
\toprule
    % \multirow{2}*{\multicolumn{2}{c}{model}} & \\  $\texttt{eval}_{\texttt{man}}$ & $\texttt{eval}_{\texttt{sge}}$ &  \\
    \multicolumn{2}{c}{Modeling method \& Setup} & $\texttt{eval}_{\texttt{man}}$ & $\texttt{eval}_{\texttt{sge}}$ \\
    \midrule
    \multirow{2}*{AT+CE} & Transformer & 23.1 & 26.8  \\
        & \quad + SA & 19.6 & 23.3 \\
    \midrule
    \multirow{6}*{\shortstack{NAT+CE}} & R-Mask~\cite{MaskCTC} & 23.1 & 26.7  \\
    & C-Mask & 23.3 & 26.5\\
    & E-Mask & 23.5 & \textbf{26.1} \\
    & M-Mask & \textbf{23.1} & 26.8 \\
    & S-Mask & 23.8 & 27.0 \\
    & F-Mask & 23.6 & 26.9 \\
    % \cline{2-5}
    \midrule
    \multirow{2}*{\shortstack{NAT+MWE}} & Output-Nbest & 23.5 & 26.7 \\
    & Input-Nbest & 23.4 & 26.6\\
    \bottomrule

\end{tabular}
\end{table}

Table~\ref{tab:CS-WER-results} reports the MER results averaged among all code-switching points. 
From Table~\ref{tab:CS-WER-results},  we find the NAT method is much worse in MER at  code-swithcing points than the AT method. Besides, language specific masking method seems  favor corresponding recognition of the languages that are masked. This suggests NAT-based code-switching ASR has potential to improve.
% We observe a similar situation as we found in Tabel~\ref{tab:maskctc-results} that is the best recognition result for code-switching points in ${eval}_{sge}$ is obtained by E-Mask method and the M-Mask method achieves the best result in ${eval}_{man}$. 
% Also, we find that the better recognition results on a whole sentence do not correspond to the better results only on those code-switching points, as we can see the results of MWE criterion 

\section{Conclusion}~\label{sec:conclusion}
In this paper, we proposed to employ Minimum Word Error (MWE)  criterion to train non-autoregressive Transformer models under code-switching ASR scenario. To obtain a strong non-autoregressive baseline for code-switching ASR, we attempted different token masking methods at code-switching points in utterance. We found the proposed masking methods can yield improved performance. 
Based on the best CE trained NAT model, the  MWE criterion further yields mixed error rate reduction, thanks to the proposed Nbest-list-based hypothesis generation methods. Since our Nbest hypothesis generation methods are a little bit mismatched with what we are doing for masking during decoding, the performance might be compromised. In future, we are exploring more desired Nbest list generation methods, hopefully yielding more gains in terms of MER reduction.

\vfill\pagebreak
\clearpage

\bibliographystyle{IEEEbib}
\bibliography{strings,refs}

\end{document}